\documentclass[twocolappendix,twocolumn]{aastex631}
\usepackage{graphicx}
\usepackage{mathptmx}
\usepackage{textcomp}
\usepackage{latexsym}
\usepackage{multirow}
\usepackage{natbib}
\usepackage{float}
\usepackage{hyperref}
\usepackage{xcolor}
\usepackage{gensymb}
\usepackage{longtable}
\newcommand{\orcid}[1]{\href{https://orcid.org/#1}{\textcolor[HTML]{A6CE39}{\includegraphics{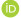}}}}
\providecommand{\keywords}[1]
{
  \small	
  \textbf{\textit{Keywords---}} #1
}
\newcommand{\ltsim}{\raisebox{-.5ex}{$\;\stackrel{<}{\sim}\;$}}
\newcommand{\gtsim}{\raisebox{-.5ex}{$\;\stackrel{>}{\sim}\;$}}
\newcommand{\kms}{\ifmmode {\rm km\ s}^{-1} \else km s$^{-1}$\fi}

\newcommand{\msun}{$M_{\odot}$}

\newcommand{\mbhdot}{$\dot{M}_{\rm BH}$}
\newcommand{\lsf}{$L_{\rm SF}$}
\newcommand{\lagn}{$L_{\rm AGN}$}

\newcommand{\hst}{HST}
\newcommand{\her}{Herschel}
\newcommand{\ltir}{$L_{\textrm{\tiny TIR}}$}

\defcitealias{T11}{T11}
\defcitealias{T17}{T17}
\defcitealias{M12}{M12}
\defcitealias{N14}{N14}
\defcitealias{N20}{N20}
\defcitealias{drizzlepac_handbook}{Drizzlepac Data Handbook}

\newcommand{\One}{SDSS J033119.66$-$074143.1}
\newcommand{\Two}{SDSS J134134.19$+$014157.7}
\newcommand{\Three}{SDSS J151155.98$+$040802.9}
\newcommand{\Four}{SDSS J092303.52$+$024739.6} 
\newcommand{\Five}{SDSS J132853.65$-$022441.6}
\newcommand{\Six}{SDSS J093508.49$+$080114.5}

\newcommand{\sOne}{\hbox{SDSS J0331--0741}}
\newcommand{\sTwo}{\hbox{SDSS J1341+0141}}
\newcommand{\sThree}{\hbox{SDSS J1511+0408}}
\newcommand{\sFour}{\hbox{SDSS J0923+0247}} 
\newcommand{\sFive}{\hbox{SDSS J1328--0224}}
\newcommand{\sSix}{\hbox{SDSS J0935+0801}}

\graphicspath{{./}}

\shortauthors{Thomas et al.}

\begin{document}

\title{Searching for the Role of Mergers in Fast and Early SMBH Growth:\\ Morphological Decomposition of Quasars and Their Hosts at $z\sim4.8$}

\author[0000-0002-2456-3209]{Marcus~O.~Thomas}
\author[0000-0003-4327-1460]{Ohad~Shemmer}
\affil{University of North Texas, 1155 Union Cir Denton, TX 76201 USA;  {\rm \href{mailto:marcusowenthomas@gmail.com}{MarcusOwenThomas@gmail.com}}}
\author[0000-0002-3683-7297]{Benny~Trakhtenbrot}
\affil{School of Physics and Astronomy, Tel Aviv University, Tel Aviv 69978, Israel}
\author[0000-0003-1523-9164]{Paulina~Lira}
\affiliation{Departamento de Astronomia, Universidad de Chile, Camino del Observatorio 1515, Santiago, Chile}
\author[0000-0002-6766-0260]{Hagai~Netzer}
\affil{School of Physics and Astronomy, Tel Aviv University, Tel Aviv 69978, Israel}
\author[0000-0001-5882-3323]{Brooke~D.~Simmons}
\affil{Department of Physics, Lancaster University, Lancaster LA1 4YB, UK}
\author{Neta Ilan}
\affil{Department of Condensed Matter Physics, Weizmann Institute of Science, 76100 Rehovot, Israel }

\begin{abstract}
	
\noindent We present rest-frame ultraviolet (UV) images of six luminous quasars at $z\sim4.8$ obtained with the Hubble Space Telescope (\hst).
These quasars exhibit a wide range of star formation rates (SFRs) and lie in a wide range of environments.
We carefully model and subtract the point-like quasar emission and investigate the morphology of the underlying host galaxies at kpc scales.
The residual images allowed identification of potential companion sources, which enabled us to explore the role of galaxy merger scenarios in the co-evolution of the quasars and their hosts.
We also search for the mechanism driving extreme SFRs in three of the quasars.
We find that the rate of detection of potential companions to the host galaxies does not follow trends between high- and low-SFR sources; i.e., the HST imaging suggests that both high- and low-SFR sources are found in both dense and sparse galactic environments.
The suggested role of major mergers driving extreme SFRs cannot be supported by the multiwavelength data in hand.
Three of four companion sources, previously revealed by sub-millimeter observations, are not detected in the HST images of three of our quasars.
An adapted high-resolution imaging strategy focused on high-SFR sources and extended to a larger quasar sample is required to determine the role of mergers in the processes of star formation and supermassive black hole growth at high redshift.
%%%
%%%
%%%

\end{abstract}

\section{Introduction}
\label{sec:intro}
The evolutionary path of quasars and the interactions with their host galaxies is not yet well-established.
Observations have shown that the supermassive black holes (SMBHs) in the centers of these quasars experience periods of rapid growth that appears to peak at $z\gtsim2$ \citep[see, e.g.,][]{miyaji01,hasinger05,richards06,silverman08,croom09,arid08,arid10,arid15,shen20_L_func}. 
Additionally, several studies over the last two decades have shown that these phases of rapid growth exhibit potential correlations with the host galaxies, particularly, a potential relationship between the hosts' star formation rates (SFRs) and SMBH mass growth (\mbhdot), probed by bolometric AGN luminosity (\lagn; see, e.g., \citealt{netzer09,lutz10,shao2010,rosario2013,shen2020,stemo2020}, but also \citealt{pagenat2012,woo2020}).  
%%%

%%%
Two main evolutionary paths regarding the SFR--\mbhdot\ correlation have emerged from observations of nearby sources, each depending on whether the system's emission is dominated by the active galactic nucleus (AGN) or star formation (SF).
In systems that are dominated by SF luminosity (\lsf), the host SFRs and AGN growth appear uncorrelated \citep[e.g.,][]{shao2010, H2010}.
However, in AGN-dominated systems, a clear trend has been observed that can be represented by \hbox{$L_{\rm SF}({\rm erg\ s}^{-1}) \simeq 10^{43}(L_{\rm AGN}/10^{43} {\rm erg\ s}^{-1})^{0.7}$} (e.g., \citealt{netzer07}, \citeyear{netzer09}; \citealt{lutz08}).

This observed link suggests that in quasars with rapidly growing SMBHs, which also exhibit extreme SFRs  (e.g., \hbox{SFRs $\gtsim10^3$ \msun\ yr$^{-1}$} for AGNs with \hbox{\lagn$\sim10^{47}{\rm erg\ s}^{-1}$}), the AGN and SF could both be fueled by common gas flow mechanisms on galactic scales. 
%%%
While not yet well-established, it has been long suggested that the observed co-evolution, and indeed the source of in-flowing material, is related to major mergers between gas-rich galaxies \cite[e.g.,][]{sanders88,dimatteo05,hopkins06,somerville08,Hopkins12,Rowlands15,Pawlik18,Lanz22}.
Gathering observational evidence to test this merger-driven co-evolutionary scenario has been difficult as most of the optical/UV and near-infrared (NIR) spectral bands are contaminated by AGN emission \citep[e.g.,][]{vandenberk2001}.
This is particularly troublesome for identifying and characterizing UV-bright star-forming galaxies in the cosmic neighborhoods of higher-redshift quasars.
Methods have been developed and refined over the years to model and remove the point-like emission produced by an AGN and reveal the out-shined structure of the underlying host galaxy, out to $z\sim3$ \citep[see, e.g., ][]{boyce98,dunlop03,kim08,kim17,simmons08,peng10,krist11,simmons12,glikman15,stark18,ghosh20,lupo2020}.
%%%%

%%%%
However, implementing these methods on higher-redshift sources requires deep dedicated observations that face potential technical challenges for optical and NIR instruments, i.e., producing high-resolution imaging with both a stable point-spread function (PSF) as well as dynamic range sufficient to prevent saturation.
Such observations are nonetheless essential to understanding the relationship between SMBH growth and the mechanisms fueling extreme SFRs, and, in particular, the role that interacting sources play in host+AGN evolution.
%%%%

Recently, such investigations have been performed on a sample of quasars at \hbox{$z\simeq4.8$}, presented in \citet[][hereafter T11]{T11}, that also has multiwavelength observations in \citet{M12}, \citet{N14}, \citet{T17}, and \citet{N20}; hereafter, M12, N14, T17, and N20, respectively. 
The \citetalias{T11} sample consists of 40 luminous, unobscured quasars in the redshift range $z\sim$4.65--4.92, selected for Mg~\textsc{ii}~$\lambda2798$ emission-line measurements through ground-based NIR spectroscopy. 
This emission line enabled
%studied in \citetalias{T11} using the VLT/SINFONI and Gemini-North/NIRI instruments, which provided
estimates of $M_{\textsc{bh}}$ and mass-weighted accretion rates (i.e., Eddington ratios; $L/L_{\text{Edd}}$).
The initial results showed that these sources are powered by SMBHs with typical $M_{\textsc{bh}}\simeq8\times10^8 M_{\odot}$ and $L/L_{\text{Edd}}\simeq0.6$, and represent the epoch of fastest growth for the most massive BHs.
\citetalias{M12} and \citetalias{N14} performed followup studies using Herschel/SPIRE and Spitzer campaigns in the mid- and far-infrared (FIR) to allow determination of the SFRs in the host galaxies.
These studies revealed SFRs in the range of \hbox{$\sim$1000--4000$~M_{\odot}~\text{yr}^{-1}$} for one quarter of the sample and $\sim400~M_{\odot}~\text{yr}^{-1}$ for the remainder of the sources \citep[through stacking analysis; see][]{Netzer2016}.
Further, and most notably, all of these sources show remarkably uniform SMBH properties (in terms of $L_{\text{AGN}}\text{ and } M_{\textsc{bh}}$) regardless of their FIR brightness.
%%%%%

The limited spatial resolution of the \her\ observations did not allow probing of the galaxy-scale, (sub-)arcsecond structures, or the larger-scale environments in search of a mechanism driving the extreme SFRs in the sample, initially assumed to be associated with major-mergers. 
Band-7 observations with the Atacama Large Millimeter/submillimeter Array (ALMA) obtained by \citetalias{T17} and \citetalias{N20} for an 18-source subset of the \citetalias{T11} sample were intended to probe the FIR continuum emission of the host galaxies and search for evidence of interacting companion sources or merger activity.
Interacting companion sources were detected in five of the 18 sources, one interacting with an extreme-SFR source, and four with moderate-SFR sources, suggesting that merger activity is not necessarily driving the extreme SFRs in the host galaxies.

%%%%%

To further probe the galaxy-scale structure in these sources in search for evidence of merger activity (e.g., interacting companions, tidal tails, clumped morphology) in a subset of the \citetalias{T11} sources, we use the Wide Field Camera 3 Infrared Detector \citep[WFC3/IR;][]{wfc3} onboard the Hubble Space Telescope (\hst) to obtain dedicated deep rest-frame UV imaging.
We use these data to search for potentially interacting companion sources and perform morphological decomposition to reveal host galaxy features hidden by AGN emission.  
%%%

%%%
This paper is organized as follows. 
Section \ref{sec:data_reduction} details the subset of the \citetalias{T11} sources used in \citetalias{T17} and this work, the basic SMBH properties, how the SFRs were determined for the sources,  the \hst\ observations, the reduction of data, as well as the process of morphological decomposition.
Section \ref{sec:analysis_and_dis} discusses the structure of the newly resolved galactic features, photometry preformed on newly detected companion sources and those detected by ALMA, and suggests adaptations to observation planning for future studies of like kind. 
Section \ref{sec:summary} provides a summary and concluding remarks on the sample and analysis method.
Complete source names appear in Tables and Figures while abbreviated source names appear throughout the text. 
Luminosity distances were computed using the standard cosmological model \citep[\hbox{$\Omega_{\Lambda}=0.7$}, \hbox{$\Omega_{\rm M}=0.3$}, and \hbox{$H_0=70$~\kms~Mpc$^{-1}$}; e.g.,][]{spergel}, which results in an angular scale of $\sim6.4$~kpc~arcsec$^{-1}$ at $z\simeq4.8$. 

%%%%

\begin{deluxetable*}{llccccc} %qso properties
\tablecolumns{7}
\tablecaption{Source Physical Properties}
\label{tab:qso_prop}
\tablehead{
\colhead{Sub-sample}&
\colhead{Quasar} &
\colhead{$\log{M_{\rm BH}}$\tablenotemark{\small a}} &
\colhead{}&
\colhead{$\log{L_{\rm AGN}}$\tablenotemark{\small a}} &
\colhead{$\log{L_{\rm SF}}$\tablenotemark{\small b}} &
\colhead{SFR\tablenotemark{\small b}}\\
\colhead{}&
\colhead{}&
\colhead{(\msun)}&
\colhead{$\log{L/L_{\rm Edd}}$\tablenotemark{\small a}}&
\colhead{(erg~s$^{-1}$)}&
\colhead{($L_{\odot}$)}&
\colhead{(\msun\ yr$^{-1}$)}
}
\startdata
FIR-Bright
&\object{SDSS J$033119.66-074143.1$}
 & 8.83 & 0.08 & 47.09 & 12.85\tablenotemark{\tiny *}  & 715\tablenotemark{\tiny *} \\	
&\object{SDSS J$134134.19+014157.7$} & 9.82 & $-0.74$ & 47.26 & 13.48 & 3035\\
&\object{SDSS J$151155.98+040802.9$} & 8.42 & $-0.26$  & 46.86 &  13.23 & 1696\\
\cline{1-7}
FIR-Faint&
\object{SDSS J$092303.52+024739.6$}  & 8.68 &$-0.18$ &46.67 & 12.69 & 488 \\%
&\object{SDSS J$093508.49+080114.5$} & 8.82 &$-0.13$ &46.87 & 12.42 & 277 \\
&\object{SDSS J$132853.65-022441.6$} & 9.08 &$-0.45$ &46.81 & 12.44 & 161 \\
\enddata
\tablenotetext{a}{Obtained from \citetalias{T11}. $L_{\text{AGN}}$ is the bolometric luminosity of the AGN derived from \textit{H}-band spectroscopy of the Mg~\textsc{ii} line.}
\tablenotetext{b}{Obtained from \citetalias{T17}; based on gray-body SEDs ($T_{\text{d}}=47$~K and $\beta = 1.6$) of ALMA data.}
\tablenotetext{*}{For this source, consistency with the \her\ data and \citetalias{N14} suggest a significantly hotter grey-body SED with $T_{\text{d}}=65$~K, resulting in $\log{L_{\rm TIR}}=13.35~L_{\odot}$ and SFR$=2225$~\msun~yr$^{-1}$ (see Section \ref{sec:sfr}).}
\end{deluxetable*}

%%%%%

\begin{deluxetable*}{llclccc}[ht] %obs log
\tablecolumns{7}
\tablecaption{\hst\ Observation Log}
\label{tab:obs_log}
\tablehead{
\colhead{Sub-sample}&
\colhead{Quasar} &
\colhead{$z$} &
\colhead{Obs. Date} &
\colhead{Obs. ID} &
\colhead{Exp. Time (s)} &
\colhead{Rest-Frame Bandwidth (\AA)}
}
\startdata
FIR-Bright&\object{SDSS J$033119.66-074143.1$}
& 4.71 		& 2018 Jan 24 	& IDHQ01010 &	2391	& 2102--2802\\
&& \nodata	& 2018 Feb 20 	& IDHQ07010 &	2391 & \nodata\\
&\object{SDSS J$134134.19+014157.7$}
& 4.67		& 2017 Dec 19 	& IDHQ02010 &	2391 & 2116--2822\\
& & \nodata	& 2017 Dec 21 	& IDHQ08010 &	2391	& \nodata\\
&\object{SDSS J$151155.98+040802.9$}
& 4.69 		& 2017 Dec 29 	& IDHQ03010 &	2394 & 2109--2812\\
&& \nodata	& 2017 Dec 29 	& IDHQ09010 &	2394 & \nodata \\	
\cline{1-7}
FIR-Faint&\object{SDSS J$092303.52+024739.6$}	
& 4.66 		& 2017 Nov 1  	& IDHQ10010 &	2393 & 2120--2827\\
&& \nodata	& 2017 Nov 4  	& IDHQ04010 &	2095 & \nodata\\
&\object{SDSS J$093508.49+080114.5$}	
& 4.70 		& 2018 Feb 20 	& IDHQ12010 &	2393 & 2105--2807\\
&& \nodata	& 2018 Mar 1  	& IDHQ06010&	2393 & \nodata\\
&\object{SDSS J$132853.65-022441.6$}	
& 4.62		& 2017 Dec 16 	& IDHQ05010 &	2393  & 2135--2847\\
&& \nodata	& 2017 Dec 20 	& IDHQ11010 &	2393 & \nodata\\
\enddata
\tablecomments{All observations were obtained with the WFC3/IR detector and F140W filter ($\lambda_{\text{eff}}=1.4~\micron$)}
\end{deluxetable*}

\section{Observations and Data Analysis}
\label{sec:data_reduction}

\subsection{Sample Selection and Properties}

Our sample consists of six sources from \citetalias{T11} and \citetalias{T17}, three of which are FIR-bright (FIR-B) and three that are FIR-faint (FIR-F), as determined from their \her\ fluxes, in the redshift range $z=4.62-4.71$.
The physical properties of the sources are reported in Table \ref{tab:qso_prop} (see \citetalias{T11} and \citetalias{T17} for additional details about these sources).

\subsection{FIR Emission and SFRs}
\label{sec:sfr}
The SFRs used to classify the sources were reported in \citetalias{N14} based on \her\ data and were significantly improved in \citetalias{T17} using ALMA-informed FIR spectral energy distributions (SEDs) to estimate total IR luminosity. 
The FIR SEDs presented in \citetalias{T17} were constructed from the ALMA data in two ways: (1) by using a fixed grey-body SED with $\beta = 1.6$ and $T_{\rm d} = 47~$K \citep[e.g.,][]{willott15,ven2016}, and (2) using templates provided by \citet{chary01}.
The total IR luminosity $L_\textrm{\scriptsize TIR}(8-1000~\micron)$ obtained from the SEDs is then used to calculate SFRs such that \hbox{$\textrm{SFR}/$\msun~yr$^{-1}=L_{\textrm{\scriptsize TIR}}/10^{10}L_{\odot}$}, following the initial mass function from \cite{chab03}.
The $L_\textrm{\scriptsize TIR}$ and SFR values obtained from both methods are consistent with one another and with those reported in \citetalias{N14}, with the single exception described below.
%%%

%%%
 In the case of the FIR-B source \sOne, the ALMA informed grey-body SED resulted in an \hbox{SFR~$\simeq700$~\msun~yr$^{-1}$}, $\sim1/3$ that of the \her\ result.
In terms of the SFR distribution in the \citetalias{T11} sample, such a value would place the source directly in the unoccupied center of the gap between the FIR-F and FIR-B subsamples.
However, the discrepancy can be accounted for in the SED models used for the ALMA data; i.e., when the grey-body model is allowed to increase the dust temperature \hbox{(from $T_{\rm d}=47~\text{K~to~} 65$~K)}, the ALMA and \citetalias{N14} \her\ results are in generally good agreement (see Section 3.3.1 of \citetalias{T17} for further details).
Physically, this could suggest that radiation from the AGN is heating gas in the host galaxy (outside of star-forming regions; see, e.g., \citealt{schneider15}).
Despite this discrepancy, we consider \sOne\ an FIR-B source for consistency with \citetalias{T17}.
Given the consistency in the two methods to determine SFRs, we adopt the $L_\textrm{\scriptsize TIR}$ and SFR values resulting from grey-body SEDs ($T_{\rm d}=47~\text{K}$) in what follows, which also allows for a direct comparison with studies of quasars at $z\gtsim5$.

%%%%
%

\subsection{HST Observations and the WFC3/IR PSF}

Using the HST WFC3/IR, we observed the $z\sim4.8$ quasars from Table \ref{tab:qso_prop} (hereafter HST sources) in the rest-frame UV over two orbits, per source, with the F140W filter \hbox{($\lambda_{\rm eff}=1.4~\micron$)}.
The observation log is reported in Table \ref{tab:obs_log}.

While unmatched in its spatial resolution, the WFC3/IR detector suffers from an under-sampled PSF.
Such a PSF can compromise point-source analysis (e.g., morphological decomposition); however, there are methods to retrieve much of the information lost due to the detector design. 
The photon response across the IR detector varies on sub-pixel scales, and $\sim0.3\%$ of the pixels have completely lost their response capability \citep{anderson2016}.
Information lost due to such pixels can be recovered by dithering observations, i.e., rather than one long exposure, obtaining many shorter exposures, each with very slightly adjusted pointing \citep[e.g., ][]{dithering10}.
These short ($\sim300$s) flat-field corrected exposures (\texttt{flt}) can be ``drizzled'' together to mimic the sampling and resolution of the foregone long exposure with accurate, pixel-by-pixel weighting, as well as subtract background noise, and reject cosmic rays.
The drizzled product rigorously preserves flux, and in most cases significantly increases spatial resolution \citep{drizzlepac_handbook}. 

\hst\ is also subject to spacecraft jitter and breathing, i.e., the optical path of the mirror changes with thermal conditions and adjusts focus, meaning that the PSF of a stationary source can vary over time \citep[e.g., ][]{acs_isr18}.
Considering each of these effects, the classical method of PSF fitting with a known or standardized PSF is not immediately applicable to precise WFC3/IR photometry. 
Therefore, as we explain below, we built an effective PSF (ePSF) by modeling the profiles of stars in the fields of the individual \texttt{flt} exposures \citep{anderson00,anderson2016}.
With this ePSF in hand, we scaled and subtracted its profile from the quasar point sources in each \texttt{flt} image, and drizzled the final products for analysis.
%%%

\subsection{Constructing and Testing the ePSF}\label{sec:construct}
%%%
Constructing an ePSF without priors is an inherently degenerate problem, i.e., accurately identifying the centroid of an under-sampled star requires a pre-defined PSF to provide missing information.
Without centroid precision beyond integer pixel values, a phase bias emerges contingent on the position within each pixel with which the photons interact. 
Therefore, the profile of a star whose centroid falls on the corner of a pixel, or on the boundary between two pixels, may be significantly different from that of a star centered on a pixel.
%%%

%%%
This degeneracy, however, can be broken by dithering observations as explained above.
The adjusted pointing of the dithers in an observation produces samplings of individual stars at different locations within different pixels. 
Using this strategy, a super-sampled ePSF can be iteratively built by averaging positions and extents of stellar profiles on the detector into an effective reference frame.
With this super-sampled ePSF, we can overcome pixel-phase and profile biases across the detector.
We detail our dithering strategy and ePSF build process below.
%%%%

%%%%
Each observation in our program used a 4-point box dither pattern with between 15 and 24 exposures, per source, over both orbits.
To correct for positional offsets between dithered exposures, each source's \texttt{flt} exposures were aligned with respect to the World Coordinate System with the \texttt{Drizzlepac} module \texttt{Tweakreg} \citep{drizzlepac}.%, which builds local catalogs of sources to improve coordinate accuracy.

The images were visually inspected for candidate stars to build the ePSF, and their 2D profiles were evaluated to confirm the stars were unsaturated or uncontaminated by nearby emission.
To test the spatial and sampling dependencies of the PSF across the detector, we constructed ePSFs using various combinations of stars, both near and far from the region of interest (ROI), and those with poorer or sharper sampling than the quasar.
Using the \texttt{photutils} \citep[v1.5.0;][]{photutils} \texttt{Python} package, each \texttt{flt} image was background subtracted (via a sigma-clipped median), and sets of candidate stars were combined to create four-times supersampled ePSFs (i.e., four-times the spatial resolution of the detector).
Following the iterative prescription in \cite{anderson2016}, our ePSF model was normalized such that its sum over the inner 5.5 pixels is unity; i.e., the flux is normalized over a circular region, around the ePSF centroid, with $r=5.5~{\rm pix}$\footnote{During construction, normalization radius of 5.5~pix represents the variable portion of the PSF and allows for 0.25-pixel positional variations in the subsequent iteration.} at each step in the construction process.
%
%%%%

\begin{figure*}
	\epsscale{1.}
	\plotone{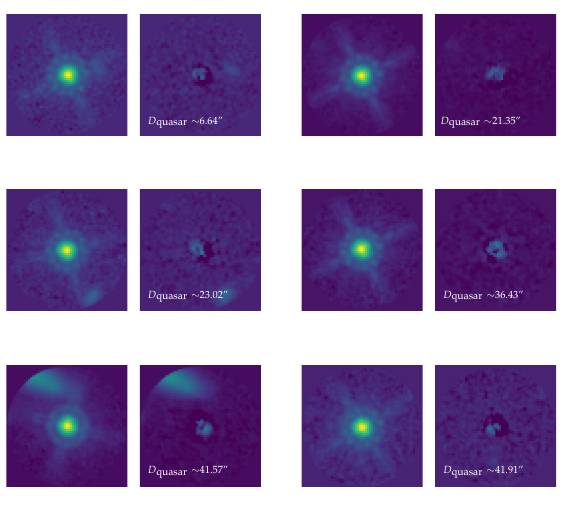}
	\caption{Cutouts of masked apertures ($r=25$~pix) of stars used to test the effectiveness of our ePSF model.
	For each pair of cutouts, the left panel is the star and the right panel is the residual following subtraction. 
The $D$ values represent the star's (projected) distance from the quasar in the image's respective ROI.
}
	\label{fig:starsub}
\end{figure*}

%%%%
Once several potential ePSF models were constructed, we evaluated their performance by fitting and subtracting them from six non-saturated test stars across the six fields, and drizzled the residual images (see Section \ref{subtraction}).
These test stars were selected such that they are (1) not severely contaminated by flux from a nearby source, (2) similar to, or lower, in flux compared with our quasars, (3) not near the edge of the field, and (4) bright enough to show significant diffraction spikes.
Our choice of test stars was motivated by allowing testing of (1) the influence of distortion across the detector, (2) whether the ePSF is too narrow/wide, and (3) how the fitter will treat diffraction spikes overlapping with other sources.
These test stars and their drizzled residuals are presented in Figure \ref{fig:starsub} and the photometric statistics are reported in Table \ref{tab:psf_field_stars}.
%%%

%%%
To quantify the quality of the ePSF fits to the test stars, we used a metric similar to that used in \citet{anderson2016,anderson22}. 
This metric ($Q)$ represents the fractional absolute value of the residuals over the inner $5\times5$ pixels as a percentage, such that 
\begin{equation}\label{eq:1}
	Q \equiv \sum|P_{\textrm{\scriptsize ij}} - s_{\textrm{\scriptsize sky}} - z_{\textrm{\scriptsize star}}\psi_{\textrm{\scriptsize ij}}| / z_{\textrm{\scriptsize star}} \times 100,
\end{equation}
where $P_{\textrm{\scriptsize ij}}$ is the flux at pixel $(i, j)$, $s_{\textrm{\scriptsize sky}}$ is the sky value, $z_{\textrm{\scriptsize star}}$ is the star's flux, and $\psi_{\textrm{\scriptsize ij}}$ is the fraction of the star's flux at pixel $(i, j)$ predicted by the ePSF model. 
We choose this metric rather than a more traditional statistic, such as $\chi^2$, as it is less sensitive to the brightness of the source, i.e., the metric quantifies departure from the ePSF shape without additional bias from pixel value weighting. 
Values of \hbox{$Q<15.0$} represent well-fit stars \citep{anderson22}.
Most of the six test stars produce $Q$-values of $\sim3$ with the highest being $\sim12$.
%%%

\begin{figure*}
	\epsscale{1}
	\plotone{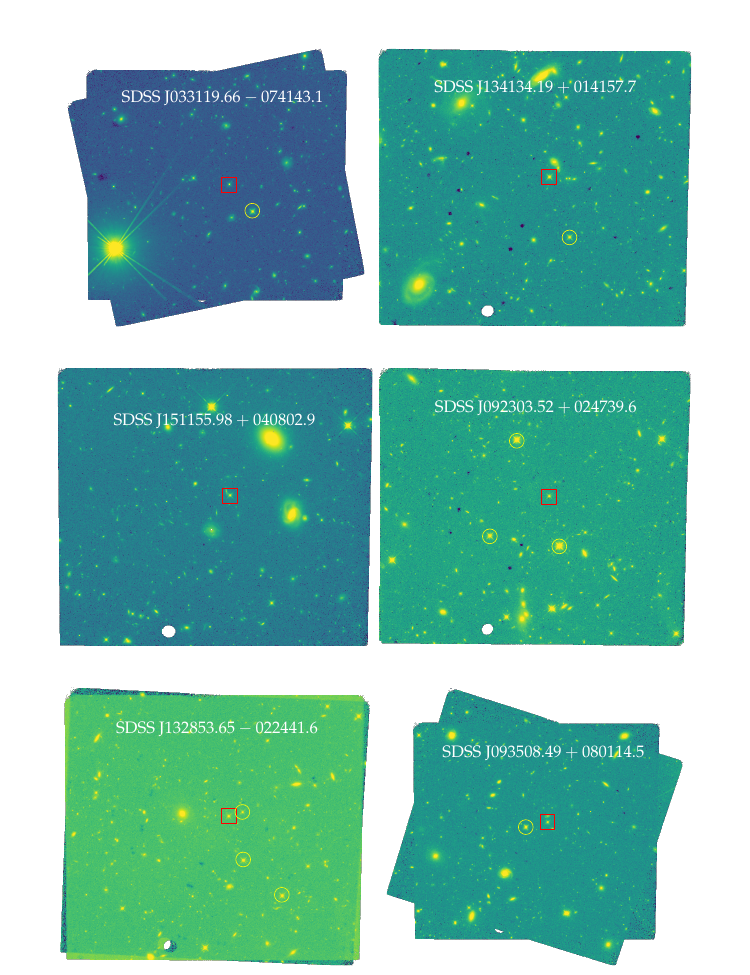}
	\caption{Drizzled images (WFC3/IR $1.4~\micron$) of the six sources in the sample. Image scaling is log-based.
	Boxes represent the quasars and circles represent stars used in the ePSF.}%
	\label{fig:psf_field_stars}
\end{figure*}

%%%
The final WFC3/IR F140W ePSF used in the quasar subtraction of our six sources is composed of nine individual stars (consisting of 157 \texttt{flt} samplings).
The drizzled images of the quasar fields and the nine stars used in the ePSF are presented in Figure \ref{fig:psf_field_stars}.
Of these nine stars, two were used as part of the six test stars (see Figure \ref{fig:starsub}), given the limited availability of stars in the quasar fields, hence a modest level of degeneracy is inherent to our ePSF construction.
The final ePSF model is presented in Figure \ref{fig:ePSF}.
We provide additional discussion on the ePSF model in Appendix \ref{apx:psf}, including the choice of stars, smoothing, as well as sources of error.

\begin{deluxetable}{lccr}
\label{tab:psf_field_stars}
\caption{Field Star Residuals}
\tablehead{
%\colhead{QSO Field}&
\colhead{$D_{\text{quasar}}$}&
\colhead{Sum}&
\colhead{Mean}&
%\colhead{$\sigma$}&
\colhead{$Q$}\tablenotemark{\footnotesize \dag}}
\startdata
6.7\arcsec & 320.1		/ -0.5 & 0.6 / -0.0 & 2.9 \\
21.4\arcsec & 1010.2	/ 10.9 & 1.9 / 0.0 	& 2.5 \\
23.0\arcsec & 293.0		/ 3.5 & 0.6 / 0.0 	& 11.8 \\
36.4\arcsec & 685.7		/ 12.9 & 1.3 / 0.0 	& 4.3 \\
41.6\arcsec & 754.3 	/ 5.4 & 1.4 / 0.0 	& 3.6 \\
41.9\arcsec & 321.4 	/ -1.6 & 0.6 / -0.0 & 3.7 \enddata
\tablecomments{Sum and mean columns are those of the pixel values in the circular ($r=11$~pix) region around each star in the image and are formatted as star/residual.\\
}
\tablenotetext{\dag}{Quality-of-fit metric; $Q<15$ represents well-fit stars \citep{anderson2016,anderson22}. See Section Eq. \ref{eq:1} and \ref{sec:construct}.}
\end{deluxetable}

\subsection{ePSF subtraction and Drizzling}
\label{subtraction}
Once a reliable ePSF was constructed, each of the \hbox{(\textit{non}-background subtracted)} \texttt{flt} exposures was background subtracted using the \texttt{MMMBackground} class, which uses the DAOPhot MMM mode-estimator algorithm \citep{daophot} where \hbox{$\text{BG}=3\times \text{median}- 2\times \text{mean}$}.
PSF photometry and subtraction were preformed on each background-subtracted \texttt{flt} image using the \hbox{\texttt{BasicPSFPhotometry}} class. 
Each source's centroid was found by fitting a 2D quadratic to the $5\times5$-pixel region around its estimated center pixel with an accuracy threshold of 0.001 pixels. 
The uncertainties were derived from the covariance matrices produced during the fits, and, in all cases, are $<1\%$.
The ePSF model flux was then scaled to the flux of the quasar's inner $5\times5$ pixels and subtracted from the \texttt{flt} image.

After the ePSF was subtracted from the quasar in each \texttt{flt} image, the individual exposures were drizzled together with the \texttt{DrizzlePac} module \texttt{AstroDrizzle}.
In a study such as this, where the exceptionally weak PSF-subtracted residuals are of prime scientific interest, optimizing the drizzle step for maximal resolution while mitigating excess noise is of utmost importance. 
Great care was taken when selecting parameters used in the \texttt{AstroDrizzle} process, and all observations were drizzled with a pixel fraction of 0.9 (i.e., each \texttt{flt} pixel was scaled to 90\% of its original size to provide finer sampling) and a final pixel scale of \hbox{0.0645 arcsec pix$^{-1}$}, or one half of the native WFC3/IR scale. 
Finer sampling parameters than those we used appear to introduce additional noise identified by visual inspection.
Final sky subtraction was achieved with the sky-subtraction step of the \texttt{AstroDrizzle} process using the ``localmin'' method.

Figure \ref{fig:qsosub} presents the quasars and their drizzled residual images.
The statistics for each quasar and their residuals are reported in Table \ref{tab:qso_stats}.
The residuals of the drizzled images were convolved with a Gaussian kernel ($\sigma=1.5,\ \text{size}=3$~pix) and the resulting smoothed images are presented in Figure \ref{fig:smoothed}.
%Residual images in the above figures are presented in linear scaling to emphasize subtle features in the inner structure of the host. 
Additionally, Figure \ref{fig:zscale} presents the residual images  scaled with the zscale algorithm\footnote{\url{https://js9.si.edu/js9/plugins/help/scalecontrols.html}}; 
 this scaling emphasizes pixel values near the image median, which, for our images, significantly increases contrast between background and faint features on the outer regions of the sources.

\begin{deluxetable}{lccr}
\label{tab:qso_stats}
\caption{Quasar - ePSF Statistics}
\tablehead{
%\multicolumn{5}{c}{First \texttt{flt} Image}\\
%\cline{2-5}
\colhead{Quasar}&
\colhead{Sum}&
\colhead{Mean}&
%\colhead{$\sigma$}&
\colhead{\textit{R}\tablenotemark{\footnotesize \dag}} 
}
\startdata
\One & 1010.3 / 22.0 & 1.9 / 0.0 & 2.2 \\
\Two & 1309.2 / 32.3 & 2.5 / 0.1 & 2.5 \\
\Three & 795.3 / 18.7 & 1.5 / 0.0 & 2.4 \\
\Four & 383.6 / 4.8 & 0.7 / 0.0 & 1.2\\
\Six & 651.4 / 16.1 & 1.2 / 0.0 & 2.5\\
\Five & 533.0 / 7.7 & 1.0 / 0.0 & 1.4 \\
\enddata
\tablecomments{Sum and mean columns are those of the pixel values in the circular ($r=11$~pix) region around each quasar in the image and are formatted as quasar/residual.\\
}
\tablenotetext{\dag}{Percent of flux remaining as a residual ($100$~times the ratio between the second and first values in Column 2).}
\end{deluxetable}

\begin{figure}
	\epsscale{1}
	\plotone{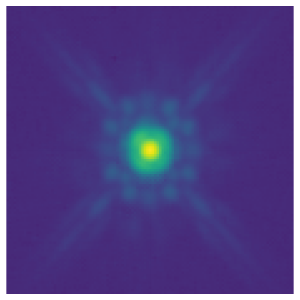}
	\caption{Four-times oversampled WFC3/IR F140W ePSF constructed from 157 isolated star cutouts in the \texttt{flt} exposures. Image is log-scaled.}
	\label{fig:ePSF}
\end{figure}

\begin{figure*}
	\epsscale{1.1}
	\plotone{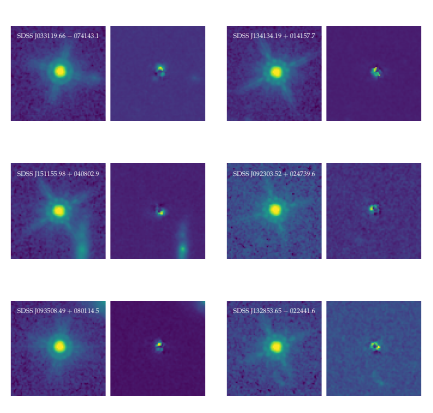}
	\caption{In each pair of $5\arcsec\times5\arcsec$ drizzled cutouts, the left panel represents the quasar and the right panel represents the residual following ePSF subtraction. Images are log-scaled.}
	\label{fig:qsosub}
\end{figure*}

\section{Analysis and Discussion}
\label{sec:analysis_and_dis}

\subsection{Source Segmentation and Photometry}

Using the image segmentation tools in \texttt{photutils}, each subtracted drizzled image was smoothed with a Gaussian-like kernel, and a source catalog was constructed ($5\sigma$ detection threshold) from the segmentation images.
Overlapping sources were de-blended if the magnitude difference between their local maxima was $\geq7.5$.
Once the detected nearby sources were cataloged, photometry was performed in two ways, both of which are explained below.

\subsubsection{Photometric Uncertainty}
With the drizzled nature of the images, calculating uncertainties must be handled as delicately as combining the original \texttt{flt} images.
The original \texttt{flt} images are provided with an error array (in one of the FITS extensions), which details the gain-based Poisson noise and photometric uncertainty due to quantum efficiency on a per-pixel basis. 
However, the physical meaning of these error arrays is lost in the drizzle process.
Since we are analyzing such faint sources and features, an inverse-variance map (IVM) weighting method was used in the final drizzle step.
IVM weighting uses various random variables (e.g., local sky, detector read noise) at each pixel to minimize the variance of the weighted average error.
The resultant data product allows greater reliability when working with very faint or small sources and features, but does not produce an accurate representation of Poisson noise in the weight map generated in the drizzle process.

To create the most accurate weight map possible, the drizzle process was repeated on the \texttt{flt} images with the science image arrays replaced with the error arrays (see Section~5.2.10 in the Drizzlepac Handbook; \citealt{drizzlepac_handbook}). 
This process produced an IVM-weighted drizzled error array using the same process and parameters applied in the drizzled science images. 
Photometric uncertainties reported hereafter are at the $1\sigma$ level, and include both detector read noise and Poisson noise in the background.
%%%

%%%
We also consider the uncertainty produced by the ePSF model construction. 
\cite{anderson2016} robustly showed that the pixel-phase bias of the detector is not significant beyond the inner $\sim6\times6$ pixels, and photometric uncertainties were shown to vary significantly less than astrometric uncertainties (see their Figure 5). 
However, as we are subtracting flux in this process, there is the potential that removing the quasar light can bring an already faint companion below our detection threshold of $5~\sigma$. 
%

%%
%%%

%%%
To evaluate the impact of the ePSF uncertainty beyond \hbox{$r=5.5~$pix}, we consider those cases where the extent of the quasar PSF overlaps with a potential companion source's profile. 
In particular, we consider the two potential companion sources that lie closest to their respective quasar (see Section \ref{sec:indiv_sources}). 
In both cases, we computed two metrics for comparison: (1) the background sky values in the respective regions for each of the sources' \texttt{flt} images and (2) the pixel values in the scaled ePSF which overlapped with the potential companion source.
In all images, the values of the overlapping ePSF pixels are significantly smaller than the background sky ($\ll 0.1\%$), and, therefore, the fluxes of the potential companions are not significantly affected by ePSF uncertainty. 
All other source detections lie further from their respective quasar compared with those two that we tested.
 Moreover, all other potential companions lie outside of the fitting region of the ePSF model.
Therefore, none of the potential companions are significantly affected by ePSF uncertainty. 

Considering the average behavior of background in our observations, as well as the flux of the scaled ePSFs used for subtraction, we find that the residual pixel values beyond $r=5.5~$pix ($r=11$ in the drizzled images) from the centroids of the test stars and the quasars are consistent with the background.
Therefore, we considered any source detection beyond this inner region to be greater than the $5\sigma$ level.

\subsubsection{Residual Source Photometry}
\label{sec:resid_phot}

Flux measurements and photometric errors were computed by integrating the total flux in a source's detected pixel footprint in the segmentation image.
Since a portion of the flux from the features in question may have been removed in the ePSF subtraction process, we used Kron aperture photometry \citep{kron} on the sources as well, which assumes a galactic nature of the residual and companion sources. 
Kron photometry estimates an elliptical isophotal region containing the majority of a source's flux, regardless of the source's galactic profile (i.e., S\'ersic index). 
This is performed by calculating the first `image moment', which represents a luminosity-weighted radius, called the Kron Radius, $R_1$.
Integrating the flux out to $2.5~R_1$ includes over $90\%$ of a source's total flux.
However, see \citet{graham_driver_2005} for a review of arguments that Kron flux estimates using $2.5~R_1$ can be missing up to $\sim50\%$ of a source's light when the respective S\'ersic index is $\gtrsim1$. Our flux estimates show no significant change when integrating out to $4~R_1$, and we therefore do not consider there is a significant loss of flux resulting from the choice of an $R_1$ factor of 2.5.
Additionally, UV luminosities at rest-frame 2500~\AA\ for all detections, assuming the same redshift as the respective quasar, was extrapolated from the source's flux according to $F_{\lambda}\propto \lambda^{-1.56}$ \citep{vandenberk2001}.
This assumption allows us to test whether the luminosities of these sources diverge from what would have been expected were they physical companions of the quasars.
Detections and properties of sources within a projected distance $\sim50~$kpc from the quasars and host galaxy residuals are reported in Table \ref{tab:comp_sources}.
The drizzled, ePSF-subtracted, and segmentation images of the $20\arcsec\times20\arcsec$ quasar fields with the Kron apertures plotted over detections of potential companion sources are presented in Figures \ref{fig:fir_b_companions} and \ref{fig:fir_f_companions} for the FIR-B and FIR-F subsamples, respectively.

We have also constructed composite images of the residuals and the contours derived from the \citetalias{T17} ALMA [C~\textsc{ii}] and continuum emission maps.
These images are presented in Figure \ref{fig:alma}.
The \texttt{flt} residual images were re-drizzled to match ALMA's pixel scale of 0.06\arcsec~pix$^{-1}$.
Both the HST and ALMA data in Figure \ref{fig:alma} are aligned to the sources' Sloan Digital Sky Survey \cite[SDSS; ][]{sdss2020} \textit{r}-band positions.
At this finer level of sampling, it was necessary to set the pixel fraction to 1.0 to ensure that every output pixel received input from the \texttt{flt} pixels.
In these frames, the HST-ALMA positional uncertainty is $\ltsim0.06\arcsec$. 
These images are used in the analysis and discussion below.

\begin{deluxetable*}{lccccccc}
\tabletypesize{\fontsize{9}{9.}\selectfont}
\tablecaption{Properties of Detected Potential Companions}
\tablehead{
\colhead{Quasar Field}&
\colhead{$D_{\text{quasar}}$}&
\colhead{$F_\textrm{k}$\tablenotemark{\tiny \ a}}&
\colhead{$F_\textrm{s}$\tablenotemark{\tiny \ a}}&
\colhead{$\log{L_{\textrm{\scriptsize UV,k}}}$\tablenotemark{\tiny b}}&
\colhead{$\log{L_{\textrm{\scriptsize UV,s}}}$\tablenotemark{\tiny b}}&
\colhead{Projected Area}&
\colhead{\hspace{-0.15cm}Color\tablenotemark{\tiny \ c}}\\
\colhead{}&
\colhead{(kpc; projected)}&
\multicolumn{2}{c}{$(10^{-20}~\textrm{erg~cm}^{-2}~\textrm{s}^{-1}~\textrm{\AA}^{-1})$}&
\multicolumn{2}{c}{(erg~s$^{-1}$)}&
\colhead{(kpc$^2$; projected)}&
\colhead{\hspace{-0.15cm}}
}
\startdata
\object{SDSS J$033119.66-074143.1$} & 1.45 & 29.08$\pm$1.24 & 29.25$\pm$1.14 & 44.20 & 44.20 & 23.00 & b- \\
\nodata & 2.43 & 2.24$\pm$0.23 & 2.06$\pm$0.21 & 43.08 & 43.05 & 4.09 & r- \\
\nodata & 12.70 & 6.32$\pm$0.68 & 3.58$\pm$0.27 & 43.53 & 43.29 & 9.20 & g- \\
\nodata & 45.43 & 14.21$\pm$1.03 & 6.87$\pm$0.37 & 43.88 & 43.57 & 17.55 & w- \\
%%%%
\object{SDSS J$134134.19+014157.7$} & 0.52 & 47.94$\pm$1.33 & 41.99$\pm$1.16 & 44.41 & 44.35 & 23.86 & b- \\
\nodata & 32.88 & 29.85$\pm$1.07 & 18.45$\pm$0.44 & 44.21 & 44.00 & 25.90 & r- \\
\nodata & 32.99 & 81.96$\pm$1.82 & 44.08$\pm$0.65 & 44.64 & 44.37 & 53.85 & g- \\
\nodata & 33.41 & 15.86$\pm$0.95 & 8.98$\pm$0.34 & 43.93 & 43.68 & 15.00 & w- \\
\nodata & 33.67 & 4.20$\pm$0.48 & 1.08$\pm$0.13 & 43.35 & 42.76 & 2.22 & c- \\
\nodata & 34.28 & 7.03$\pm$0.69 & 2.25$\pm$0.20 & 43.58 & 43.08 & 4.77 & m- \\
%\nodata & 35.34 & 333.12$\pm$1.53 & 303.16$\pm$1.18 & 45.25 & 45.21 & 181.14 & y- \\
%%%%
\object{SDSS J$151155.98+040802.9$} & 1.05 & 28.64$\pm$1.35 & 23.88$\pm$1.09 & 44.19 & 44.11 & 21.13 & b- \\
\nodata & 10.82 & 44.48$\pm$1.46 & 21.72$\pm$0.50 & 44.38 & 44.07 & 32.21 & r- \\
\nodata & 17.53 & 110.89$\pm$1.61 & 79.30$\pm$0.76 & 44.78 & 44.63 & 75.15 & g- \\
\nodata & 26.25\tablenotemark{\scriptsize*} & 14.04$\pm$0.83 & 7.50$\pm$0.32 & 43.88 & 43.61 & 12.78 & w- \\
\nodata & 31.18 & 8.43$\pm$0.68 & 4.42$\pm$0.26 & 43.66 & 43.38 & 8.52 & c- \\
\nodata & 36.43 & 7.90$\pm$0.71 & 2.01$\pm$0.20 & 43.63 & 43.03 & 4.77 & m- \\
\nodata & 38.90 & 4.25$\pm$0.47 & 2.35$\pm$0.19 & 43.36 & 43.10 & 4.60 & y- \\
\nodata & 41.16 & 6.65$\pm$0.62 & 3.19$\pm$0.22 & 43.55 & 43.23 & 6.30 & b. \\
%%%%
\object{SDSS J$092303.52+024739.6$} & 1.09 & 6.71$\pm$1.22 & 5.33$\pm$0.87 & 43.56 & 43.46 & 9.20 & b- \\
\nodata & 44.39 & 38.26$\pm$0.57 & 36.25$\pm$0.43 & 44.31 & 44.29 & 24.20 & r- \\
%%%%
\object{SDSS J$093508.49+080114.5$} & 0.60 & 25.06$\pm$1.39 & 19.83$\pm$1.15 & 44.13 & 44.03 & 22.32 & b- \\
\nodata & 19.52 & 22.07$\pm$1.25 & 10.05$\pm$0.43 & 44.07 & 43.73 & 22.49 & r- \\
\nodata & 20.76 & 6.42$\pm$0.70 & 2.04$\pm$0.21 & 43.54 & 43.04 & 5.45 & g- \\
\nodata & 22.98 & 129.52$\pm$2.17 & 89.27$\pm$1.04 & 44.84 & 44.68 & 127.63 & w- \\
\nodata & 23.86 & 31.54$\pm$1.53 & 10.70$\pm$0.47 & 44.23 & 43.76 & 26.92 & c- \\
\nodata & 28.54 & 102.47$\pm$1.62 & 87.56$\pm$0.94 & 44.74 & 44.67 & 98.83 & m- \\
\nodata & 29.08 & 15.65$\pm$1.14 & 4.41$\pm$0.33 & 43.92 & 43.37 & 13.29 & y- \\
\nodata & 47.40 & 12.34$\pm$0.65 & 9.41$\pm$0.35 & 43.82 & 43.70 & 13.80 & b. \\
\nodata & 48.68 & 11.52$\pm$0.55 & 9.90$\pm$0.33 & 43.79 & 43.73 & 12.78 & r. \\
%%%%
\object{SDSS J$132853.65-022441.6$} & 0.70 & 12.59$\pm$1.39 & 8.06$\pm$1.07 & 43.83 & 43.64 & 11.25 & b- \\
\nodata & 19.16 & 5.49$\pm$0.39 & 4.45$\pm$0.22 & 43.47 & 43.38 & 6.13 & r- \\
\nodata & 34.58 & 18.32$\pm$0.91 & 11.17$\pm$0.36 & 43.99 & 43.78 & 17.72 & g- \\
\nodata & 37.92 & 7.28$\pm$0.64 & 1.83$\pm$0.18 & 43.59 & 42.99 & 4.09 & w- \\
\nodata & 40.11 & 15.43$\pm$0.77 & 9.80$\pm$0.33 & 43.92 & 43.72 & 14.48 & c- \\
\nodata & 44.18 & 3.99$\pm$0.38 & 2.98$\pm$0.19 & 43.33 & 43.20 & 4.60 & m- \\
\nodata & 46.52 & 8.56$\pm$0.50 & 6.19$\pm$0.26 & 43.66 & 43.52 & 8.69 & y- \\
%%%%
\enddata
\tablecomments{The first row for each quasar field is the feature most likely associated with the host galaxy, selected by distance from the quasar's sky coordinates followed by visual inspection.}
\tablenotetext{a}{$F_{\text{k}}$ and $F_{\text{s}}$ represent Kron and segment flux, respectively. See section \ref{sec:resid_phot}.}
\tablenotetext{b}{UV luminosities at rest-frame 2500~\AA\ assume the sources lie at the redshifts of the respective quasars, and are extrapolated from fluxes assuming $F_{\rm \lambda}\propto\lambda^{-1.56}$ following \citet{vandenberk2001}. Kron and segment subscripts are the same as in the note above.}
\tablenotetext{c}{Color and style of Kron aperture in Figures \ref{fig:fir_b_companions} and \ref{fig:fir_f_companions}. Key: (b) blue; (r) red; (g) green; (w) white; (c) cyan; (m) magenta; (y) yellow; (-) Solid line; (.) Dotted line.}
\tablenotetext{*}{This source resides at coordinates consistent with a detected companion in \citetalias{T17}.}
\label{tab:comp_sources}
\end{deluxetable*}

%--------------
%-------------- OBJ 1
%--------------

\subsection{Notes on Individual Sources}\label{sec:indiv_sources}

Among those sources detected in the residual images, those potentially interacting with the host galaxies are discussed below.

\subsubsection*{\sOne} %obj1
The residual image of \sOne (first row of Figure \ref{fig:fir_b_companions}) reveals an extended feature in the host's southern region as well as a separate source that appears to connect to this feature.
The additional source (red aperture) falls outside of the ALMA contours in Figure \ref{fig:alma}.
%%%
We note that the extended feature connecting to the separate source traces the western side of a darker region that suggests over-subtraction by the ePSF.
This region is indeed consistent with the average position determined as the quasar centroid by the ePSF fitter.
However, the extended feature and companion source are sufficiently separated such that they fall out of the region where the ePSF exhibits significant inter-pixel variation (i.e., inner $\sim$3--5 pixels), suggesting that both of their structures are physical and that the ``empty'' region north of the companion is uncertain. 
We find that the overall shape of the host residual is generally consistent with the size and morphology suggested in the \citetalias{T17} continuum maps, i.e., the UV regions are mostly contained within the host FIR continuum.

%%%

%%%
The residual image also reveals a source 2\arcsec\ ($\sim$13~kpc, projected) to the quasar's west, which was previously hidden by the PSF wing (see green aperture).
There is not a prior detection near this new source's coordinates in the SDSS \citep{sdss2020}, SIMBAD Astronomical Database \citep{simbad}, nor in the NASA/IPAC Extragalactic Database\footnote{\url{https://ned.ipac.caltech.edu}} (hereafter, ``Databases'' collectively).
This source's flux is \hbox{$\sim$15\%} of \sOne's host galaxy flux and neither this source nor the host galaxy appears to have extended features suggesting physical interaction. 
While the ALMA data did not resolve this source to the host's west, the uniform gradient in the \citetalias{T17} \hbox{[C~\textsc{ii}]} velocity map (see their Figure 5) as well as the radial distribution in the velocity-dispersion map, suggest rotationally-dominated kinematics of an undisturbed disk.
Under these inferred kinematics, it is likely that this potential companion and the feature mentioned above are indeed projected, and a major merger scenario is not supported.
%

%--------------
%-------------- OBJ 2
%--------------

\subsubsection*{\sTwo}%obj2
The residual image for \sTwo\ contains the largest residual flux in our sample.
Such a result should be expected as it is also our highest SFR source and exhibits the highest FIR continuum flux in the sample, (\hbox{SFR$~\approx3000$~\msun~yr$^{-1}$}; \hbox{$F_{\scriptsize 152\micron}=18.5~{\rm mJy}$}), and it hosts a SMBH with a mass of $\approx6.6\times10^9$~\msun, $\sim10\%$ of its host's dynamical mass \citepalias{T17}.
However, while residing in a relatively dense field, we find no evidence for ongoing interaction with a companion.
Two extended sources are detected south east of \sTwo\, both of which we excluded from consideration as companion sources.
The larger of the two is consistent with the detection of the galaxy \hbox{SDSS J$134134.47+014151.9$} at $z\approx0.5$ \citep{obj2Comp1,obj2comp2}.
The additional source directly north of the galaxy at $z\approx0.5$ is not present in the Databases; however its size, resolved structure, and large flux are likely consistent with a closer source at a lower redshift than the quasar, and we do not consider it as a companion. %

%%%

%%%
A faint feature extending from the quasar host galaxy in the westward direction becomes visually apparent with the image scaling used in Figure \ref{fig:zscale}.
This feature is considered significant and its extent can be seen in the segmentation map in the second row of Figure \ref{fig:fir_b_companions}.
While the ALMA data suggests a relatively coherent and disk-like structure, the emission maps trace a feature consistent with the size and direction seen of the faint feature in the new \hst\ residuals (see Figure \ref{fig:alma}).
The nature of this feature is unclear, however it could suggest late stages of a major merger; i.e., coupled with relative disk-like structure, this feature could trace the structure during final coalescence. 
The available data are insufficient to determine whether the host galaxy is indeed in the later stages of a merger, however, interacting companions are not significantly detected in the ALMA data or our \hst\ images.

%--------------
%-------------- OBJ 3
%--------------

\subsubsection*{\sThree}%obj3

The field around \sThree\ is relatively dense with new detections. 
Previously, partially hidden by the PSF wing, two overlapping sources are revealed to the host galaxy's southwest.
Both of these sources are highly elongated \hbox{(Kron aperture eccentricity~$\gtsim0.8$)}.
There is not a prior detection of either of these sources in the Databases, and these sources were not detected in the ALMA data.
Considering their size and resolved structure, these are likely projected neighbors residing at lower redshifts.
%%%

%%%
The FIR continuum emission maps \citepalias{T17} revealed two faint companion sub-millimeter emitting sources (see third panel of Figure \ref{fig:alma}). 
These two sources, labeled ``SMG'' and ``B'', reside 14 and 25~kpc from \sThree, respectively.
Only SMG showed significant emission ($\sim3.5\sigma$) in the [C~\textsc{ii}] source map, and its redshift is similar to that of \sThree. 
Both B's lack of detected [C~\textsc{ii}] emission as well as evidence of the source across the multi-wavelength (HST+ALMA) data available for \sThree\ suggests that it may be a \textit{projected} nearby source. 
\citetalias{T17} did note that it is unlikely to find multiple sub-millimeter emitting sources with the observed fluxes and SFRs in a single ALMA pointing, which could indicate that source B is a true interacting companion. 
%%%%%

%%%%%
In our \hst\ imaging, a source is significantly detected ($5\sigma$) that is associated with the position of source B in the ALMA data (see white aperture in row three of Figure \ref{fig:fir_b_companions} as well as the upper right panel in Figure \ref{fig:alma}), and the region associated with SMG is consistent with background emission ($3\sigma$ upper flux limit of $2.9\times10^{-20}$~erg~s$^{-1}$~cm$^{-2}$~\AA$^{-1}$). 
We discuss the detection and non-detections as well as the nature of the ALMA companions in Section \ref{sec:role}.
The remaining sources in the quasar field do not show physical features suggesting interaction with the host galaxy.
%--------------
%-------------- OBJ 4
%--------------

\subsubsection*{\sFour} %obj4

Both \sFour's residual image and nearby field are not indicative of ongoing merger activity.
Similar to the structure suggested in the ALMA velocity dispersion maps, the host residual appears disk-like and undisturbed.
We note a previously undetected source with flux six-times that of \sFour's host is detected $6.9\arcsec$ in the southeast direction (see red aperture in first row of Fig \ref{fig:fir_f_companions}) in the \hst\ imaging.
If this source has a similar redshift to the quasar, it would reside $\sim44~\text{kpc}$ away, and could be involved in merger activity with \sFour\ as well as the \citetalias{T17} spectroscopically confirmed companion sub-millimeter galaxy (SMG) $\sim35$~kpc to the southwest (see SMG in the lower left panel of Fig. \ref{fig:alma}).
We, however, fail to detect the additional ALMA-detected interacting source to the southwest in our \hst\ imaging ($3\sigma$ upper flux limit of $1.5\times10^{-20}$~erg~s$^{-1}$~cm$^{-2}$~\AA$^{-1}$).
%%%
%--------------
%-------------- OBJ 6
%--------------

\subsubsection*{\sSix} %obj6

\begin{figure*}
%	\epsscale{0.85}
	\plotone{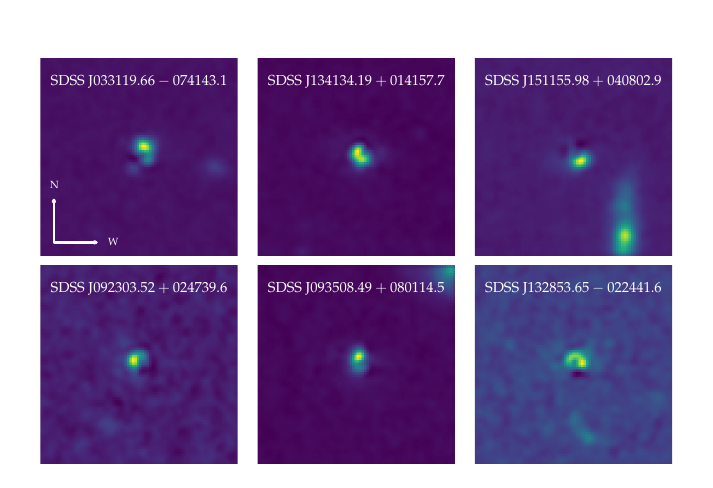}
	\caption{Gaussian-kernel ($\sigma=1.5,\ \text{size}=3$~pix) smoothed, drizzled quasar residuals (WFC3/IR $1.4~\micron$). Image scaling is linear. The compass in the first panel applies to all six images.}
	\label{fig:smoothed}
\end{figure*}

\begin{figure*}
%	\epsscale{0.85}
	\plotone{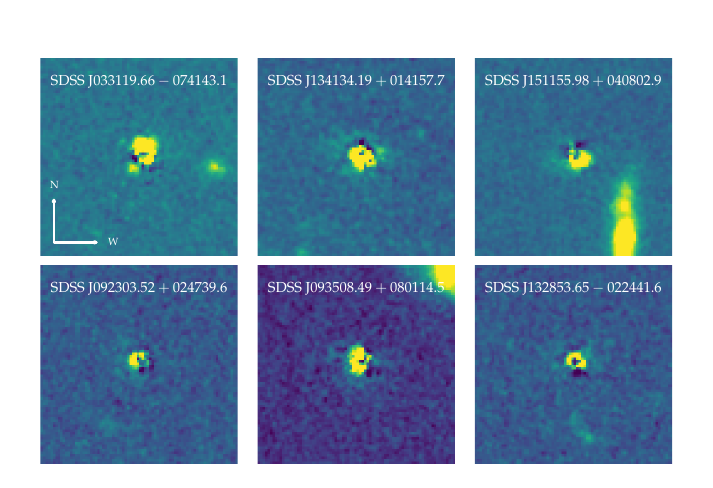}
	\caption{Residual images (WFC3/IR $1.4~\micron$) plotted with a zscale interval, a histogram-like plotting scale that emphasizes differences from the background. 
	Companions and extended features can be easily identified, however, structure and interior features are overwhelmed by binning.
	The compass in the first panel applies to all six images.}
	\label{fig:zscale}
\end{figure*}

\begin{figure*}
	\epsscale{1.2}
	\plotone{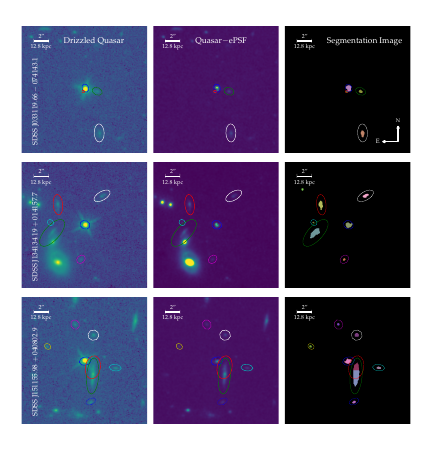}
	\caption{The leftmost column shows the original drizzled image of each FIR-B quasar. 
	The middle column shows the residual image following ePSF subtraction.
	The rightmost column shows the deblended segmentation image of each significantly detected source (i.e., the footprint in pixels of individual sources). 
	The apertures in each row represent the Kron profile of sources within $\sim50~$kpc (projected) of the quasar, assuming a typical galactic morphology (see text and Table \ref{tab:comp_sources}).
	Aperture color is consistent within each row.
	Note that image normalization varies with each row and the brightness of companions.
	The compass in the first row applies to all images.}
	\label{fig:fir_b_companions}
\end{figure*}

The residual image of \sSix\ presents an interesting case.
Extending south from the host appears a resolved non-nuclear structure on a kiloparsec scale, suggesting a disturbed morphology (see the lower center panel of Figure \ref{fig:smoothed}).
This is further supported by the \citetalias{T17} [C~\textsc{ii}] velocity and continuum emission maps.
Although it was the lowest signal-to-noise source in \citetalias{T17} ($S/N\sim4.5$), the emission maps suggest an irregular, non-rotating structure.
\citetalias{T17} did not detect a companion in this case.

Eight sources are detected within $8\arcsec$ of \sSix's host galaxy, none of which appear in the Databases.
Half of these sources appear to be clustered in a region directly north-northwest of the host galaxy.
Considering the small likelihood of multiple sources being found so close together in the redshift range of the quasar, as well as the size and resolved structure between the clustered sources, it is likely that this is a grouping of projected neighbors at a lower redshift than \sSix.
Additional observations are needed to classify the remaining detected sources as projected neighbors or physical companions.
%

%%%
%--------------
%-------------- OBJ 5
%--------------
\subsubsection*{\sFive} %obj5

\begin{figure*}
	\epsscale{1.2}
	\plotone{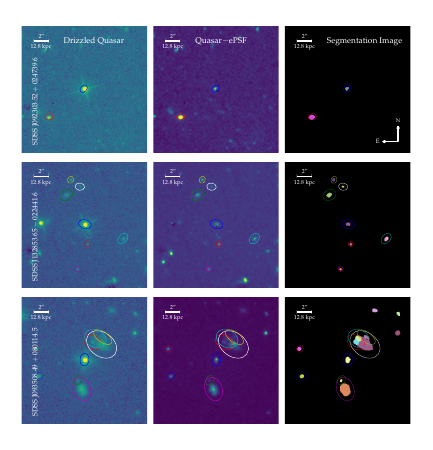}
	\caption{Same as Figure \ref{fig:fir_b_companions} but for FIR-F quasars.}
	\label{fig:fir_f_companions}
\end{figure*}

\begin{figure*}
	\epsscale{1.2}
	\plotone{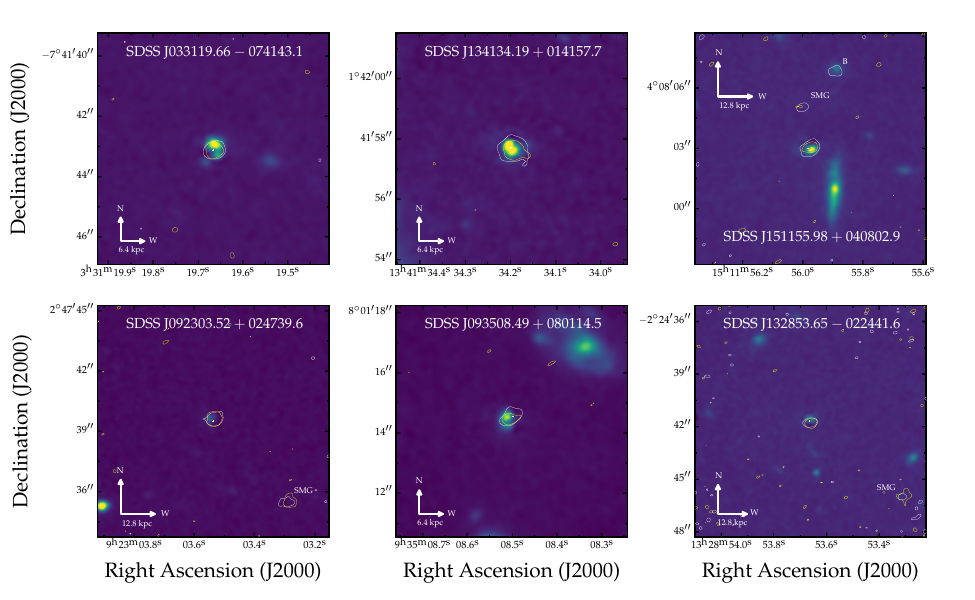}
	\caption{
	Quasar residual images (WFC3/IR $1.4~\micron$) with ALMA continuum ($\lambda152~\micron$) and [C~\textsc{ii}] ($\lambda157.74~\micron$) contours (from \citetalias{T17}) plotted in white and yellow, respectively. The contours trace the respective emission at $2.5~\sigma$ significance. The SDSS optical positions of the sources are represented by white crosses. Images with ALMA-detected companions are extended to show the respective contours. The kpc scale in each panel measures the base-to-tip length of the horizontal compass arm.}
	\label{fig:alma}
\end{figure*}

The structure of the host residual for \sFive\ is uncertain in our \hst\ imaging. 
The source appears to have been subject to significant over-subtraction in its center-most pixels.
To evaluate the performance of our ePSF subtraction in the source's observations, we included a nearby star with a similar flux to the quasar in our testing during construction of the ePSF, and we found the fitting to behave well, with residuals consistent with the background. 
The results can be seen in the upper left pair of panels in Figure \ref{fig:starsub}, as well as the lower center panel of Figure \ref{fig:fir_f_companions}. 

Considering the above, it is possible that much of the host galaxy's star-forming structure extends well beyond the nuclear region.
We do not find evidence for merger activity in the host residual itself, however, an imprint of the ePSF's central pixels is seen in the un-smoothed residual image, suggesting that some information may have been lost during subtraction.

The case for companion sources near \sFive\ is similar to that of both \sThree\ and \sFour; i.e., \citetalias{T17} spectroscopically confirmed a true companion galaxy $\sim40$~kpc to the source's southwest, yet the same location in our \hst\ imaging is consistent with background ($3\sigma$ upper flux limit of $4.7\times10^{-20}$~erg~s$^{-1}$~cm$^{-2}$~\AA$^{-1}$).
%%%



\subsection{The Role of Mergers in Our Sample}

\citetalias{T17} concluded that the FIR-F sources \sFour\ and \sFive\ were each interacting with a single SMG and the FIR-B source \sThree\ was interacting with (potentially) two companions: an SMG and a ``blob'' without significant line emission, i.e., no redshift measurement.
Our resolved host galaxies are in general agreement with the morphologies inferred from the ALMA velocity maps (i.e., the HST residuals are similar in size and generally contained within the ALMA contours).
However, we fail to detect three of the four companion sources.
The coordinates for ALMA-detected SMG companions interacting with \sThree, \sFour, and \sFive\ are consistent with the background in our \hst\ imaging, while the ``blob'' near \sThree\ is significantly detected in the \hst\ imaging.
Further, we find that disturbances in disk-like structure shown by the contours in the ALMA velocity maps are well represented in our host residuals.
%%%

%%%
Given the performance of our ePSF model on field stars as well as the consistency of disturbances with the ALMA data, we consider that our residuals contain the resolution necessary to view structure on scales larger than the latest (coalescence) stages of major galaxy mergers.
Structure \textit{within} the hosts is still uncertain given possible influence from intra- and inter-pixel variation in the WFC3/IR detector.
%%%

Our \hst\ imaging is unable to detect the three interacting sources suggested by the ALMA data. 
We do, however, find that our results support \citetalias{T17}'s suggestion that \sSix's host galaxy is likely disturbed.
The FIR-B source \sTwo\ and the FIR-F source \sSix\ exhibit potential evidence of late-stage merger activity.
However, our ePSF model is insufficient to confirm such results on the scale of these particular features.
Both the FIR-B and FIR-F sub-samples have sources residing in dense as well as sparse galactic environments.
Even if similar redshifts to the host galaxies are assumed for these newly detected sources in the dense environments, the major-merger scenario cannot necessarily be identified as the driving force behind extreme SFRs and rapid SMBH growth, as equally extreme rates are observed in sources without neighbors. 
We conclude that, in our sample, extreme SFRs are not necessarily associated with major merger activity; i.e., evidence for major merger activity is seen in some low-SFR systems and some high-SFR systems do not show evidence of interaction with companions.

%%%

\subsection{The ALMA Companions}
\label{sec:role}

The non-detection of the companion SMGs reported in \citetalias{T17} is a curious case and raises questions about the nature of SMG sources.
The nature and abundance of SMGs in the $z\gtsim4$ regime is not well defined, however they are understood to be large, extremely luminous starburst galaxies, often powered by gas-rich mergers \citep[e.g., ][]{casey14}. 
Additionally, SMGs are characterized by high levels of obscuration, which could explain their absence in our \hst\ images.
However, the rest-frame UV emission is not necessarily always hidden behind obscuring dust. 
\cite{chen15} performed an \hst\ rest-frame UV survey of 48 SMGs up to $z\sim3$, which yielded a detection rate of 80\%, suggesting that obscuration is not systematically aligned with SF regions and stellar populations.
%%%

\begin{figure*}
	\epsscale{1.1}
	\plotone{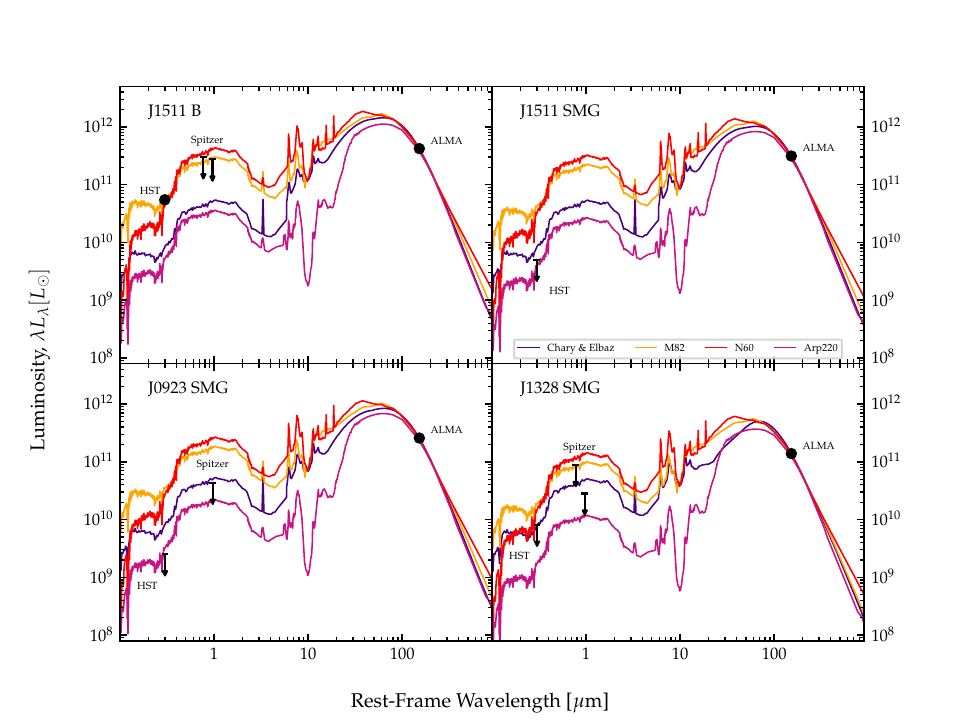}
	\caption{SEDs for the four ALMA-detected companion sources assuming different levels of star formation and dust obscuration (see Section \ref{sec:role}). 
	Each template is anchored to the respective ALMA continuum measurement (black circles; \citetalias{T17}). 
	Downward facing arrows represent the HST and Spitzer (observed-frame 3.6 and 4.5~$\micron$) upper limits. 
	Arrow length is $3\sigma$, with the downward tip representing $1\sigma$.
	The companion in the top-left panel was significantly detected by HST and is also represented by a black circle.
	Uncertainty on the ALMA, and single HST, data points are smaller than the size of the respective markers.
	}
\label{fig:sed}
\end{figure*}

%%%
To constrain the nature of these companions, Figure \ref{fig:sed} presents SEDs constructed for the four sources using templates representing different types of galaxies.
It is important to note that these SEDs represent a limited set of potential models for these companion sources.
As these companions are assumed to be at $z=4.8$, as well as in a state of merger, we chose the following templates to represent luminous, merging, and star-forming galaxy types.
The SEDs contain the M82, N60, and Arp220 templates from the SWIRE Template Library \citep{polletta}, which represent starburst, luminous IR starburst, and dust-obscured ultraluminous IR galaxy (ULIRG) types, respectively. 
Additionally, we use the \cite{chary01} templates which correspond to the respective $L_\textrm{\scriptsize TIR}$ and $\lambda152~\micron$ ALMA measurements reported in \citetalias{T17}.
%

%%%

%%
Each SED template is anchored to the source's ALMA continuum measurement.%
Also plotted are the \hbox{1--3$\sigma$} upper limits from both the new HST imaging and the \citetalias{N14} Spitzer/IRAC observations, in which these companion sources were also not detected.
These upper limits are derived from aperture photometry in the $r=2\arcsec$ region associated with the ALMA detections.
While this size aperture is smaller than the $r=3\arcsec.6$ recommended radius\footnote{\url{https://irsa.ipac.caltech.edu/data/SPITZER/docs/irac/}}, its selection was necessary to avoid emission from the nearby target quasars.
We note that the single case where the spatial resolution is insufficient for separating the SMG from the quasar is in the Spitzer image of \sThree; in this case, upper limits on the companion SMG could not be measured.
Additionally, the $3.6~\micron$ Spitzer upper limit for the SMG near \sFour\ could not be reliably measured due to detector artifacts in the region; only the $4.5~\micron$ upper limit is considered a constraint.
For each companion, new \ltir\ and SFR values were calculated for those templates that the upper limits are consistent with, and these are reported in Table \ref{tab:tir_sfr}.
%%%

%%%
In the case of ``blob'' B near \sThree, \citetalias{T17} suggested the detection could trace a faint, low-mass galaxy undergoing a \textit{minor} merger with the quasar's host. 
Such a distinction is characterized by B's lack of [C~\textsc{ii}] detection in light of the clear detections in the three confirmed interacting SMGs.
This source is significantly detected in the HST imaging, and allows an additional constraint on the SED.
Both the HST and ALMA measurements suggest that, if its redshift is similar to that of the quasar, then B is possibly consistent with a starburst galaxy; i.e., the HST measurement ($\lambda_{\rm \scriptsize rest}\sim0.25~\micron$) agrees with the SEDs of both M82 and N60, and the Spitzer upper limits suggest only the M82 SED provides a viable option of those tested.
The M82 template is based on an edge-on starburst-type irregular galaxy.
Irregular galaxies lack the spiral arm structure that is typically thought to trigger SF, and M82's starburst nature is thought to be fueled by interaction with its larger companion M81. 
The N60 template is based on NGC~6090, which is a luminous IR starburst galaxy that appears to be in intermediate stages of a gas-rich major merger. 
These two galaxy types produce similar IR SEDs, especially at the wavelengths our observations cover.
However, they diverge in the far UV, suggesting that observations at shorter wavelengths would allow tighter constraints on structure apart from B's possible starburst nature and determine the influence of merger activity.
%%%

%%%
The remaining sources appear to exhibit similar nature with respect to the relatively low $\sim0.24~\micron$ HST upper limits.
In all three SMG cases, both the Arp220 and Chary \& Elbaz templates are viable at $\sim3\sigma$, and the SMG near \sFive\ could possibly be consistent with all four templates (albeit at $>3\sigma$).
We note that all three SMG sources are consistent with the Arp220 template, which represents a dust-obscured ULIRG.
Such a galaxy may elude detection in HST observations such as ours due to the high levels of dust extinction that may obscure the relatively low-power UV emission produced by the small SFRs inferred from the IR SEDs \cite[e.g.,][]{casey14,mazz19}.
However, the paucity of data points for these sources does not allow us to reliably constrain these companions to the SEDs tested.
%%%

%%%
Using the potential SEDs inferred by the detections and upper limits, new integrated \ltir(8--1000~$\micron$) values and SFRs are reported in Table \ref{tab:tir_sfr}.
In all cases, the new \ltir\ and SFR values are broadly consistent with those from \citetalias{T17}; i.e., the differences are within $0.2~\rm{dex}$.

\begin{deluxetable*}{lcccccccc}
\tablecaption{ALMA-Detected Companion $L_{\textrm{\scriptsize TIR}}$ and SFRs}
\tablecolumns{9}
\tablehead{
\colhead{Companion}&
\colhead{   $\log{L_{\text{CE}}}$\tablenotemark{\tiny a}}&
\colhead{  $\log{L_{\text{M82}}}$\tablenotemark{\tiny b}}&
\colhead{  $\log{L_{\text{N60}}}$\tablenotemark{\tiny c}}&
\colhead{  $\log{L_{\text{Arp}}}$\tablenotemark{\tiny d}}&
\colhead{ 		SFR$_{\text{CE}}$\tablenotemark{\tiny a}}&
\colhead{		SFR$_{\text{M82}}$\tablenotemark{\tiny b}}&
\colhead{		SFR$_{\text{N60}}$\tablenotemark{\tiny c}}&
\colhead{		SFR$_{\text{Arp}}$\tablenotemark{\tiny d}}\\[-1ex]
\colhead{}&
\colhead{($L_{\odot}$)}&
\colhead{($L_{\odot}$)}&
\colhead{($L_{\odot}$)}&
\colhead{($L_{\odot}$)}&
\colhead{(\msun~yr$^{-1}$)}&
\colhead{(\msun~yr$^{-1}$)}&
\colhead{(\msun~yr$^{-1}$)}&
\colhead{(\msun~yr$^{-1}$)}
}
\startdata
J1511 B\tablenotemark{\scriptsize *}
			& 12.39	& 12.45		& 12.48 	& \nodata	& 246 	& 281 		& 301		& \nodata\\
J1511 SMG	& 12.28	& \nodata	& \nodata 	& 12.10		& 191	& \nodata	& \nodata	& 126\\
J0923 SMG	& 12.16	& \nodata	& \nodata 	& 12.02		& 144	& \nodata	& \nodata	& 104\\
J1328 SMG	& 11.86	& 11.97		& 12.00 	& 11.75		&  73	& 93		& 99		& 56 \\
\enddata
\tablenotetext{a}{Total IR luminosities, \ltir\ (8--1000~$\micron$), and SFRs from \citetalias{T17}, based on the \cite{chary01} template with $\lambda 152~\micron$ continuum luminosity best matched to the ALMA measurement.}
\tablenotetext{b}{Based on the M82 starburst template.}
\tablenotetext{c}{Based on the N60 luminous IR galaxy template.}
\tablenotetext{d}{Based on the Arp220 dust-obscured ULIRG template.}
\tablenotetext{*}{This source lacks significant [C~\textsc{ii}] emission to assume a redshift measurement similar to the quasar host.}
\label{tab:tir_sfr}
\end{deluxetable*}

\subsection{Future Implementation of the Observational Strategy}

This pilot study is aiming to explore the applicability of PSF decomposition to larger quasar samples at high redshifts, and to determine how the method should be adapted in order to investigate the role of mergers.
Ideally, deep multi-orbit rest-frame UV observations of the entire \citetalias{T11} sample (or those with ALMA data; \citetalias{N20}) would likely produce more robust results that could help determine the role of mergers in AGN+host evolution.
However, meaningful results could also be produced by adapting the imaging strategy used in this paper to devote all pointing time to only FIR-B sources to search for major merger activity.

While each orbit in the observations received sufficient flux for significant detections in the quasars' regions, only $\sim1\%$ of the flux remained after ePSF subtraction (see \hbox{Table \ref{tab:qso_stats}}) that could be used for measuring host galaxy emission.
Future imaging of the quasar host galaxies in the \citetalias{T11} sample, or other $z\gtsim4$ quasars, should be focused on the sources suspected of merger-driven activity (i.e., extreme SFRs), and should be sufficiently deep ($>2$ orbits) for increased source detection.
Selecting stars for the PSF model was a difficult process with our small dataset, and few candidates existed that were not subject to geometric distortion. 
An ePSF model could be significantly improved by the addition of dedicated observations of many non-saturating stars, similar to the dataset used in \citealt{anderson2016} (which used $\sim$10$^3$ star samplings from a single set of dithered exposures) close in time to the quasar observation. 
An improved model could help to reduce the uncertainty in the inner 3--6 pixels.

\section{Summary and Conclusions}
\label{sec:summary}

Following ALMA observations of three high- and three low-SFR quasars from the \citetalias{T11} sample, we used \hst\ imaging to resolve the stellar content and environment of the respective host galaxies.
Based on reliable results from previous studies in both simulations and observations from lower-redshift samples, we used host+AGN decomposition on these six sources to evaluate the applicability to the full \citetalias{T17} sample.
%%%

%%%
To perform the host+AGN decomposition on our sample, we extracted stars from the fields surrounding the quasars to model the behavior of the WFC3/IR PSF, which varies with time, temperature, and pixel location.
We used this PSF model to subtract the point-like quasar emission, leaving behind the emission of the host galaxy, and potentially interacting companions.
%%%

%%%
We find that with sufficient sampling, previously unidentified features and sources are significantly detected. 
Nevertheless, previous sub-mm observations detected interacting companion galaxies that were not detected in the new HST imaging.
We constructed best-match SEDs for the ALMA companion sources using templates of various types of SF galaxies.
These SEDs are constrained by the ALMA data, as well as by upper limits from the new HST images and past Spitzer observations.
We find that the SF regions of these non-detected sources are highly dust-obscured, and are mostly consistent with (ultra) luminous IR galaxies.
Initially, the extremely high SFRs in three of the sources were thought to be driven by late-stage major merger activity.
This hypothesis is not supported by our findings.
Potentially interacting companions are detected in both the high- and low-SFR environments, which may suggest a secular mechanism is responsible for the increased activity.
%
%%%

%%%
As this work serves as a pilot study, we also explore ways to adjust the imaging strategy to efficiently produce more robust results.
Improving the sub-arcsecond (galactic-scale) resolution of the host galaxy morphology could be achieved by dedicating dithered observing time to unsaturated PSF stars in the center of the field of view as close in time to the quasar observations as possible. 
The quality of the ePSF is the most important consideration in this method when operating on sources with footprints spanning only several pixels.
Observing our sample with JWST could significantly improve our ability to resolve the inner regions of these quasar host galaxies and help to constrain the mechanisms responsible for the extremely high SFRs and rapid SMBH growth at this important cosmological epoch.

\begin{acknowledgments}
This research is based on observations made with the NASA/ESA Hubble Space Telescope obtained from the Space Telescope Science Institute, which is operated by the Association of Universities for Research in Astronomy, Inc., under NASA contract NAS 5–26555. 
These observations are associated with program number \hbox{HST-GO-15118.002-A}.
BT acknowledges support from the European Research Council (ERC) under the European Union’s Horizon 2020 research and innovation program (grant agreement 950533) and from the Israel Science Foundation (grant 1849/19).
BDS acknowledges support through a UK Research and Innovation Future Leaders Fellowship [grant number MR/T044136/1].
We thank an anonymous referee for helping to improve this manuscript.
This research has made use of the NASA/IPAC Extragalactic Database, which is funded by the National Aeronautics and Space Administration and operated by the California Institute of Technology.
This research also made use of the SIMBAD database, operated at CDS, Strasbourg, France.
\end{acknowledgments}

\begin{appendix}
\restartappendixnumbering

\section{On Our Effective PSF}
\label{apx:psf}

We turn to a discussion on the nature of the ePSF in general as well as those models used in this paper. 
Many software tools have been used extensively in the literature to build PSF models on various detectors across the entire redshift regime, and there is not a ``best'' tool for creating a PSF.
Rather, each tool fits a single or a few particular use cases.
For example, \texttt{TinyTim} \citep{tinytim} simulates \hst\ point-source observations and is best suited for observation planning or evaluating variability over time, and \texttt{PSFGAN} \citep{stark18} uses a generated adversarial network to perform PSF photometry on samples with thousands of $z\ltsim0.1$ sources.
Despite there being many tools used in past studies, the motivation for our choice of \texttt{photutils} relied on a few key points.
In particular, its procedure of building PSFs is based on that of \citet{anderson00}, which provides the most robust prescription to create an effective ePSF for science use. 
%%%%

%%%%
The smoothing step during construction is crucial when working with oversampled models. 
Our choice of four-times oversampling allows for the ePSF structure to vary on sub-pixel scales, i.e., each pixel on the detector is treated as a $4\times4$ grid of pixels.
Such fine oversampling is achieved by the wealth of dithered exposures provided by multiple orbits per source. 
However, using an oversampled ePSF that varies on sub-pixel scales on the larger WFC3/IR pixels can introduce non-physical variations in the structure of the residual.
This variation is mitigated in the smoothing step of building the ePSF, which is essential as the ePSF is a continuous function.
%%%%

%%%%
We use the smoothing method described in \citet{anderson2016}, which is a two-step process.
First, each iteration during the build phase is smoothed with a $3\times3$ boxcar (i.e., average) kernel. 
The smoothed ePSF produced is then subtracted from the unsmoothed ePSF.
This residual is then convolved with a $5\times5$ kernel where the inner $3\times3$ and the outer $2\times2$ represent quartic and quadratic kernels, respectively. 
As in \citet{anderson2016}, we found this process to produce a model that was not so smooth as to fail at the sharpest points nor produce non-physical structure in the inner region; the residuals between the smoothed ePSF and the unsmoothed ePSF were effectively reduced to zero. 
%

%%%
%
The WFC3/IR detector exhibits significant geometric distortion (up to 10\%) away from the center of the field, and previous studies have constructed ``gridded'' ePSFs, where the detector is split into a $3\times3$ grid and each section has its own respective ePSF. 
Based on a star's position, fitters could weight the contribution from each of the nine ePSFs.
We attempted to create such a three-dimensional ePSF, however, our fields were too sparse and each ePSF model was severely undersampled.
Since our use case for the final ePSF model is so narrow, we were able to judiciously select which stars would best represent the PSF in the quasars' on-axis positions, where geometric distortion is at its minimum. 
While a handful of other ``clean'' stars are present in the fields, those that were not used resided in regions where geometric distortion was significant, and many were saturated.
When the stars subject to distortion were included in an ePSF model, fit residuals were relatively poor.
Ultimately, we selected stars in and around the region we termed the ``region of interest'', rather than in a particular grid position.
%%%

Our fits to test stars in Figure \ref{fig:starsub} produced generally acceptable residuals in each case, even on stars near the edge of the detector.
Most notably, there does not appear to be systematic over- or under-subtraction in the test fits, i.e., every residual does not exhibit a particular dark or bright region in the same part of the removed PSF.
The appearance of such features would suggest that the model's normalization width is too wide or narrow, which seems to not be the case in our ePSF.
%%%

%%%
There do, however, seem to be artifacts in the residuals of the host galaxies suggesting insufficient sampling of the ePSF model, the most apparent being over-subtraction.
The average background in our images is Gaussian-distributed with a mean value $<0.01$.
It should be expected that a well-modeled PSF subtracted from a star would leave behind a similar Gaussian-distributed residual resembling background, but that is not the case for our model in most trials.
Our results show regions that would benefit from improved sampling of the inner $\sim$3 pixels of the ePSF that could likely be achieved with dedicated observations of a field of many stars as close in time as possible to the quasar observations.

\end{appendix}

\clearpage

\bibliography{bibliography.bib}

\end{document}